\documentclass[12pt,preprint]{aastex}
\usepackage{natbib}
\usepackage{bibentry}
\usepackage{amssymb}
\usepackage{epsfig}
\begin{document}

\title{Spatially-resolved Chandra Imaging Spectroscopy of the classical/weak-lined T Tauri system V710 Tau}

\author{Sonali J. Shukla}
\affil{Department of Physics \& Astronomy, Vanderbilt University, Nashville, TN 37235}
\email{sonali.j.shukla@vanderbilt.edu}

\author{David A. Weintraub}
\affil{Department of Physics \& Astronomy, Vanderbilt University, Nashville, TN 37235}
\email{david.a.weintraub@vanderbilt.edu}

\and

\author{Joel H. Kastner}
\affil{Chester F. Carlson Center for Imaging Science, Rochester Institute of Technology, 54 Lomb Memorial Drive, Rochester, NY 14623}
\email{jhkpci@cis.rit.edu}

\begin{abstract}

  We present spatially-resolved X-ray observations of the binary T Tauri star system V710 Tau. Using Chandra's Advanced CCD Imaging Spectrometer (ACIS), we imaged this $3.2''$ separation binary system, consisting of a classical T Tauri star, V710 Tau N, and a weak-lined T Tauri star, V710 Tau S. The Chandra ACIS-S3 images --- obtained in two 9 ks exposures separated by about three months (2004 December and 2005 April) --- cleanly resolve the V710 Tau binary, demonstrating that both stars emit X-rays and thereby enabling the first spectral/temporal study of the individual components of this mixed (classical and weak-lined) T Tauri star binary system. The northern component, V710 Tau N, appears to have been in a flaring state during the first (2004 December) exposure. During this flare event, the X-ray flux of the classical T Tauri star hardened significantly. Single-component plasma models with plasma temperatures in the range $kT_X \sim 0.7-1.1$ keV are adequate to fit the observed X-ray spectra of V710 Tau S in 2004 December and both stars in 2005 April. The 2004 December flare-state observation of V710 Tau N requires a higher-temperature plasma component ($kT_X\sim2.5 - 3.0$ keV) in addition to the soft component ($kT_X\sim0.5$ keV) and is better fit by a model that includes a slightly enhanced Ne/Fe abundance ratio. These results are generally consistent with statistical contrasts between the X-ray emission properties of classical (rapidly accreting) vs. weak-lined (weakly accreting or non-accreting) T Tauri stars.

\end{abstract}


\section{Introduction}

Young stars, and low-mass pre-main sequence (T Tauri) stars in
particular, are well-established as strong X-ray sources. Almost
certainly, the production of hard (E $\geq$ 1 keV) X-rays in the
environments of young stars is linked to magnetic reconnection events.
The stimulus of the magnetic reconnection events that generate the
X-rays, however, is still a subject of considerable debate (see
discussions in \citet{pre05c} and  \citet{tel07b}). X-ray production in active stars can be attributed to many processes. In young stars, X-rays are thought to be produced by the heating and confinement of coronal plasma by a strong
magnetic field. However, multiple processes may contribute to coronal
heating. In young stars, both dynamo activity and star-disk
interactions can, in principle, produce X-rays, but it remains to be
established whether both processes contribute to the observed X-ray
emission. If accretion of material from the surrounding disk onto the
star produces a distinct X-ray signature, X-ray emission
characteristics of classical T Tauri stars (cTTS) and weak-lined T
Tauri stars (wTTS) should differ, since cTTS are thought to be
actively accreting material from their disks while wTTS are not actively accreting.

High-resolution X-ray spectra for the classical T Tauri stars TW Hya
\citep{kas02,ste04}, BP Tau \citep{sch05}, and V4046 Sgr \citep{gue06}
indicate that the cooler X-ray emitting plasma from these stars is
produced via accretion-generated shocks. \citet{rob06} find that
accretion shocks contribute to the observed X-ray emission of several
cTTS (BP Tau, CR Cha, SU Aur and TW Hya) though the contribution
varies from source to source.  Long-term X-ray and infrared light curves obtained from observations of V1647 Ori during its accretion-powered optical/IR eruption \citep{kas06} also favor accretion as the ultimate source of
magnetically-derived X-ray emission. On the other hand, statistical
studies showing that the mean X-ray luminosity of cTTS is two to three
times lower than that of wTTS \citep{fla03,pre05b,fei05,gue06,tel07b}
--- and that rapid rotation and strong X-ray emission are positively
correlated for main sequence stars \citep{ran00} and pre-main sequence
stars \citep{ste01,sta04} --- point to pre-main sequence stellar
dynamo activity as the source of the X-rays. However, \citet{fei03}
found that pre-main sequence stellar rotation and X-ray emission were
slightly anti-correlated. Indeed, accretion may play a role in the apparent suppression of the X-ray luminosities of cTTS relative to wTTS. \citet{tel07b} suggest that this effect may be due to cooling of cTTS coronal material by accretion streams.
  
One of the more interesting lines of evidence in this debate is that presented in a study by \citet{kon01} of multiple T Tauri stars in Taurus. They found statistical evidence in ROSAT data that primaries are more X-ray luminous and produce harder X-rays than secondaries. They also found that, in cases for which rotational velocities and bolometric luminosities were known, primaries are more rapidly rotating and/or more luminous. Stronger X-ray emission, they concluded, is due either to higher bolometric luminosity or faster rotation.

Spatially-resolved X-ray observations of stars in wTTS/cTTS
binary systems have the potential to be especially valuable for
determining differences in the X-ray properties of these two classes
of pre-main sequence stars, as the component stars in such systems are
likely to have similar ages. Such observations have been rendered
feasible by the subarcsecond spatial resolution of the Chandra X-ray
Observatory. In the first example of such an observation, the borderline wTTS/cTTS system Hen3-600 was studied by \citet{hue07} with the Chandra High Energy Transmission Grating Spectrograph (HETGS). Signatures of
  accretion were found in these X-ray gratings spectroscopy data. The
  results --- though somewhat ambiguous --- implicate the more rapidly
  accreting component in the Hen 3-600 binary as the source of these
  signatures.

In this paper, we present the results of Chandra imaging spectroscopy observations of the wTTS/cTTS binary system V710 Tau. The N and S stellar components of this binary system (hereafter V710 Tau N and V710 Tau S) have been classified \citep{whi01} as an M0.5 cTTS (EW(H$\alpha$)=$-$69 \AA) and an M2 wTTS (EW(H$\alpha$)=$-$7.2 \AA)\footnote{\citet{har94} found EW(H$\alpha$)=$-$11 \AA for V710 Tau S.}, respectively. The northern component is slower rotating \citep{kon01} and is, by a small margin, the primary; \citet{whi01} derive masses of 0.68 M$_\odot$ and 0.62 M$_\odot$ while \citet{jen03} find masses of 0.68 M$_\odot$ and 0.48 M$_\odot$ for the northern and southern components respectively. However, the primary displays a bolometric luminosity ((log L$_{bol}$/L$_\odot$ = $-$0.29) twenty percent smaller than that of the secondary (log L$_{bol}$/L$_\odot$ = $-$0.20) \citep{whi01}. Hence, the available data suggest that V710 Tau N, though perhaps more evolved, is more rapidly accreting than V710 Tau S. Indeed, the contrast between the H$\alpha$ EWs of the components of V710 Tau is more pronounced than is the case for Hen 3-600 (\citet{hue07} and references therein).

The V710 Tau system has been detected as an X-ray source previously by Einstein \citep{fei81}, ROSAT \citep{neu95}, and most recently in three XMM-Newton studies \citep{fav03,giar06,gud07}. In contrast to these investigations, our Chandra data provide the first X-ray observations in which the individual components of this 3.17'' binary are spatially resolved. Thus, this is the first time in which X-ray photometry and spectral analysis has been performed on the individual components of this wTTS/cTTS binary star system. Our Chandra observations demonstrate that both stars emit X-rays and reveal the contrasting X-ray emission properties of the cTTS and the wTTS components of this system.

\section {Observations and Data Processing}

The binary T Tauri system V710 Tau was observed by Chandra on 29 December 2004 (ObsID 5425; 9043 s exposure) and again on 4 April 2005 (ObsID 5426; 8976 s exposure).  For each observation the target was positioned on the back-illuminated CCD S3 of the Advanced CCD Imaging Spectrometer (ACIS-S3).  The Chandra/ACIS-S3 back-illuminated chip, which is sensitive to X-rays in the 0.3-10 keV range, has a pixel size of 0.49" and a field of view of $\sim$8' $\times$ 8'. 

 The photon event data were processed using the standard Chandra data processing pipeline, CIAO, version 3.2. Circular spectral extraction regions of radius 1.66'' that encompass each source individually, without significant overlap, were defined. Similarly, using four circular regions, each of radius 5.0'', the background flux level centered about the binary system was defined. Since the background flux level was taken from the area surrounding the binary system, the same background regions were used for both stars in each observation. Using these regions, background-subtracted spectra were extracted using the CIAO psextract science thread\footnote{http://cxc.harvard.edu/ciao/}. The data were further processed utilizing CCD subpixel event repositioning (SER) techniques, designed to optimize Chandra/ACIS spatial resolution \citep{li03}. The improved ACIS-S3 images obtained after SER processing more clearly resolve the binary but do not affect the outermost portions of the PSFs of the two stars where the PSFs overlap. Hence, spectral analysis (section 3.2) was performed on the event data prior to SER processing.

\section{Results}

\subsection{ACIS-S3 Multiepoch Imaging and Light Curves}

\begin{figure}
\epsscale{.80}
\epsfig{file=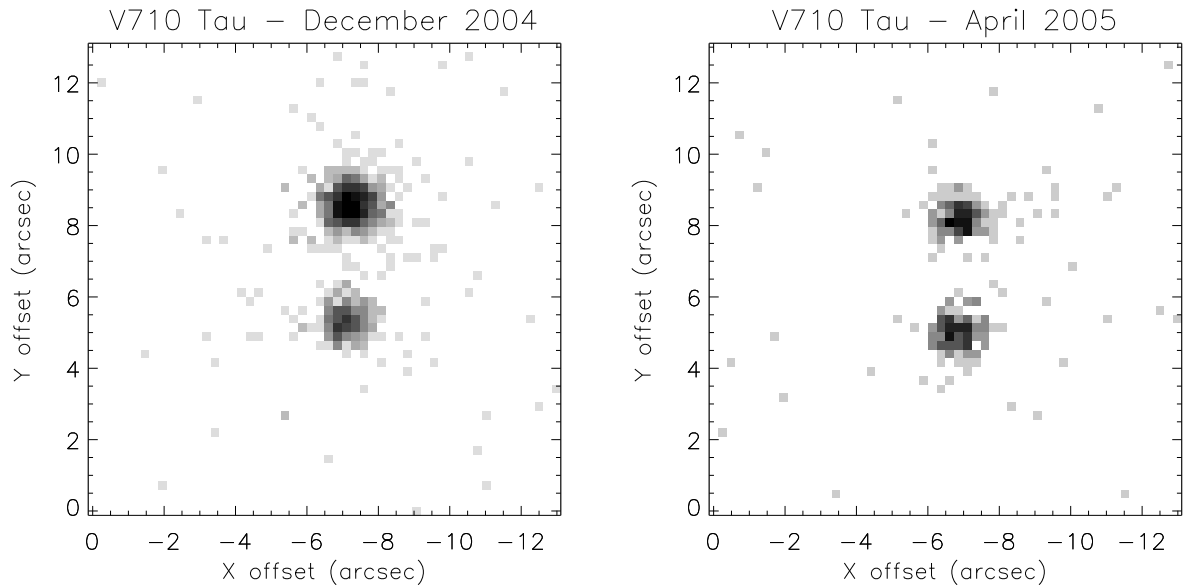,angle=90,width=6.5in}
\caption{Chandra/ACIS-S3 X-ray images of the V710 Tau binary (3.17'' separation) obtained 2004 December (left) and 2005 April (right). The grey-scale ranges of the images peak at 5.75 counts ks$^{-1}$pixel$^{-1}$ (December 2004) and 1.78 counts ks$^{-1}$pixel$^{-1}$ (April 2005). \label{fig1}}
\end{figure}

As is evident in Figure ~\ref{fig1}, the Chandra images cleanly resolve the X-ray binary system and demonstrate that both V710 Tau N and V710 Tau S are sources of X-rays. These X-ray data (see count rates in Table 1) reveal that the cTTS is the brighter X-ray source of the pair in December 2004 and that the two stars are nearly equal in X-ray brightness in April 2005.

\begin{figure}
\begin{center}
\begin{tabular}{cc}
\epsfig{file=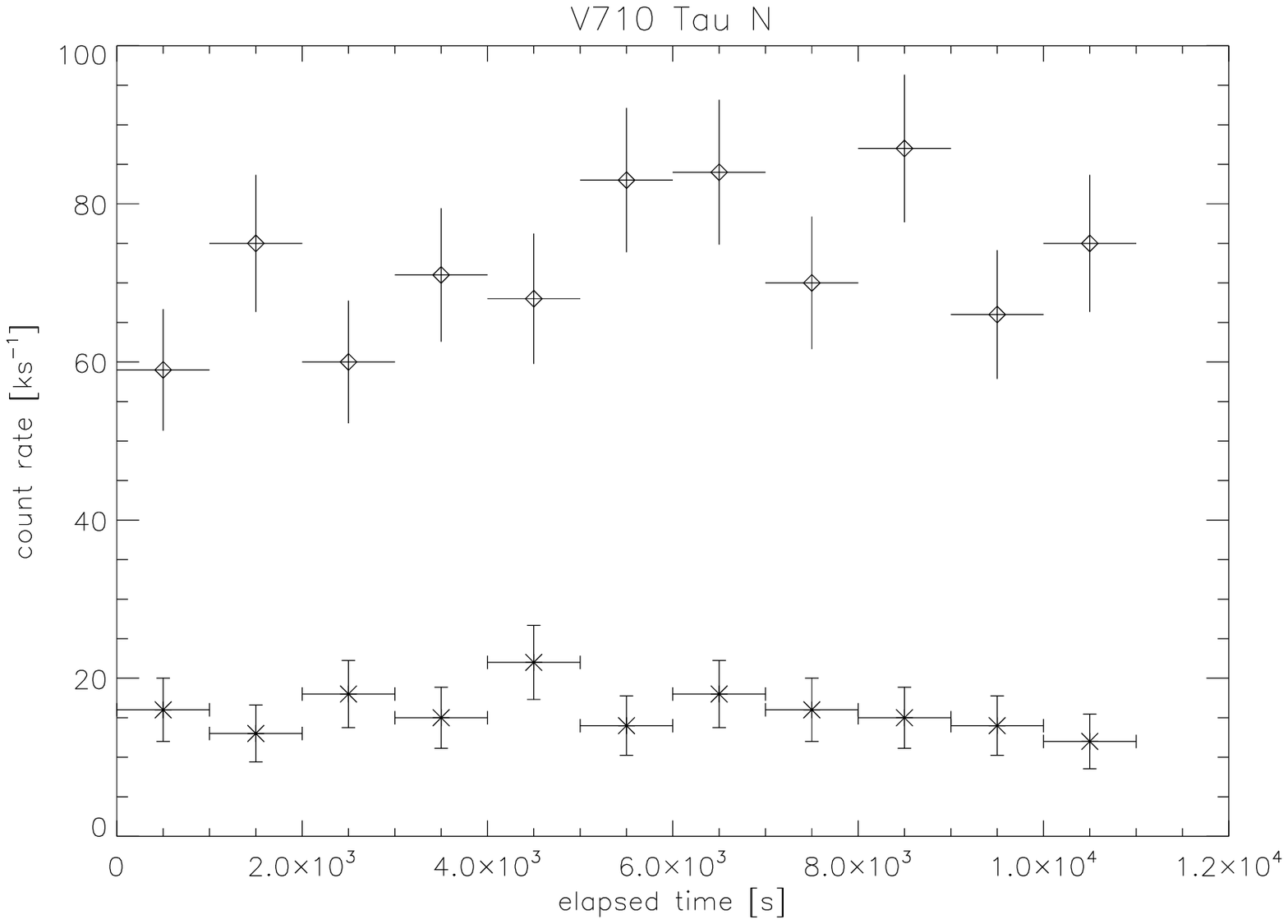,width=3.25in} &
\epsfig{file=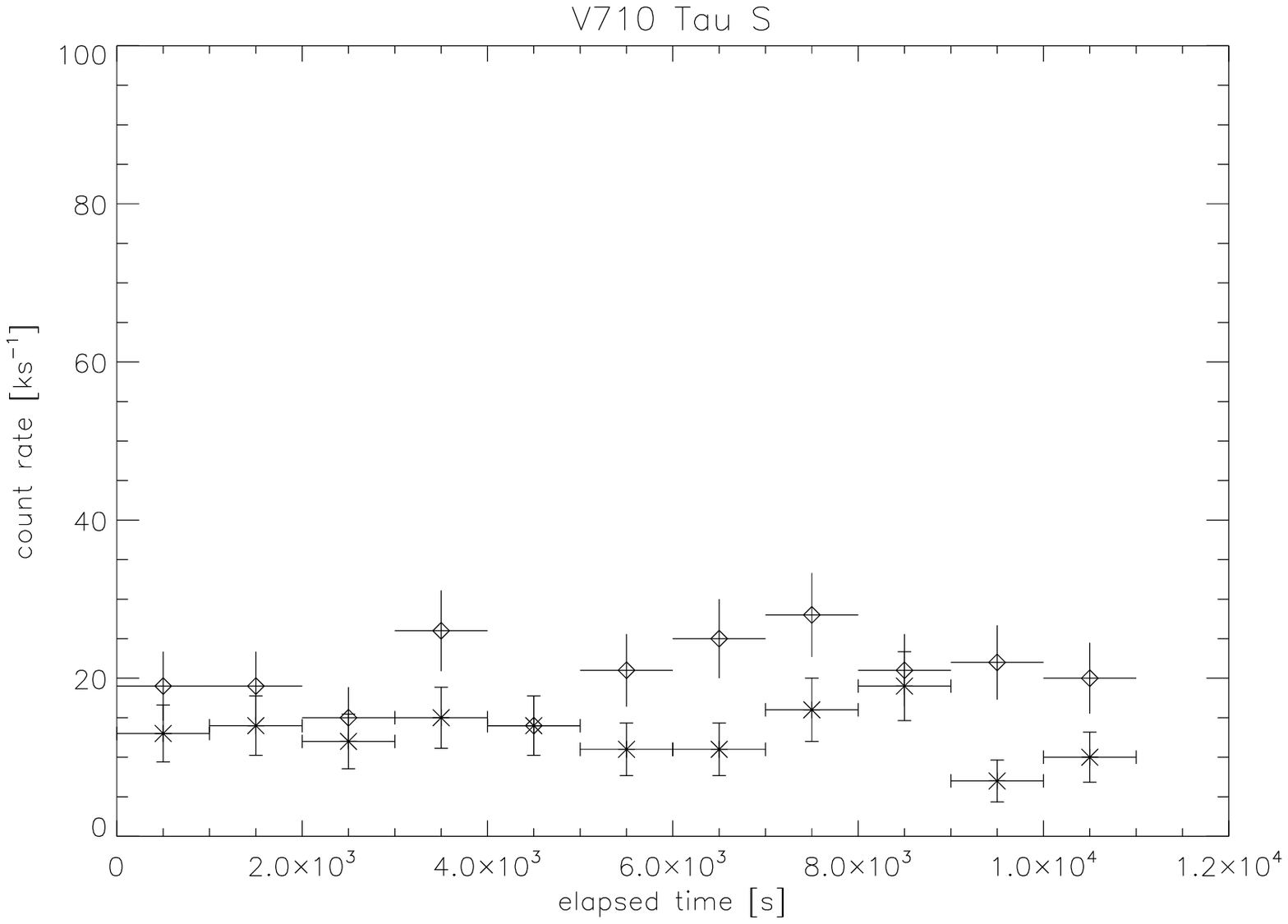,width=3.25in} \\
\end{tabular}
\caption{Light curves for V710 Tau N (left) and V710 Tau S (right) in December 2004 (diamonds) and April 2005 (crosses). Horizontal error bars indicate the width of the time bins (1000 s), and the vertical error bars indicate 1 $\sigma$ uncertainties in count rate per 1000s time interval.\label{fig2}}
\end{center}
\end{figure}

Within each 9 ks exposure, both stars appear to have nearly constant count rates (Table 1, Figure ~\ref{fig2}), although, within the 3$\sigma$ uncertainty range, the X-ray count rates may vary by as much as a factor of two on a 1 ks timescale. However, the count rates of both stars were larger in December 2004 than in April 2005.  Specifically, the average X-ray count rate of V710 Tau S was $\sim$1.5 times greater in December than in April, while the X-ray flux of V710 Tau N was $\sim$5 times larger in the first-epoch observation. In addition, the ACIS hardness ratios (HRs) for the binary (calculated according to the HR definitions in \citet{get05}) demonstrate that the emission from V710 Tau N was harder in the first observation than in the second.  In contrast, the HRs of V710 Tau S remained constant to within the uncertainties, and its X-ray emission was overall somewhat softer than that from V710 Tau N, especially in December 2004 (Table 1). These data suggest that V710 Tau N was in a flaring state during the first observation, whereas V710 Tau S displayed less significant, if any, long-term flaring.

\subsection{Spectral Analysis}

Spectral analysis was performed with the X-ray data analysis tool XSPEC, version 11.3.1 \citep{arn96}.  The spectral data in the region of 0.3 to 5.0 keV (essentially no photons were detected from 5 to 10 keV) were fit with the MEKAL code describing thermal, collisionally excited plasmas \citep{mewe85,mewe86,lie95}; the VMEKAL code, in which metal abundances can be adjusted individually; and the APEC code describing optically-thin plasmas in collisional ionization equilibrium \citep{smi01}. The intervening photoelectric absorption is incorporated through the WABS model \citep{mor83}. We find that the APEC and MEKAL model fits do not differ significantly; thus in this paper we only report the analysis performed with the MEKAL and VMEKAL models. 

For each of the two observations of each component of the V710 Tau binary (hereafter denoted as N1 and S1 for the December 2004 and N2 and S2 for the April 2005 observations of the northern and southern components, respectively), we fit the X-ray spectrum with a thermal plasma model suffering intervening absorption. \citet{whi01} measure A$_{V}$ for V710 Tau N and V710 Tau S to be 1.80 $\pm$ 0.36 and 1.82 $\pm$ 0.51 respectively.  Hence, adopting N$_{H}$ = 1.6 x 10$^{21}$ A$_{V}$ \citep{vuo03}, we use the optically-derived value of N$_{H}$ = 0.29 x 10$^{22}$ cm$^{-2}$ for our spectral modeling. Allowing N$_{H}$ to vary overconstrains the model fitting and does not produce meaningful results for the model parameters. Previous studies (e.g., \citet{giar06} and references therein) have found that plasma models with subsolar abundances better reproduce the X-ray spectra of T Tauri stars than do models that assume solar abundances. We assume a subsolar abundance of Z = 0.2 $Z_{\sun}$, which is nearly identical to the abundance value found for the combined spectrum of V710 Tau by \citet{fav03}. Indeed, we find that single-component MEKAL plasma models provide reasonable fits for all four spectra (Figs. 3-5).

For the spectrum with the best signal-to-noise ratio, obtained from observation N1, we find that a two-component model improves the spectral fit significantly in comparison to a one-component model (Table 2). A two-component model with both a cooler plasma component ($kT_{1}\sim0.5$ keV) and a hot plasma component ($kT_{2}\sim2.6$ keV) provides an improved fit, especially at energies below $\sim1$ keV (Fig. 3).  However, while such two-component models also provide adequate fits for the spectra resulting from observations N2, S1, and S2, we are unable to constrain the model normalizations, and thus the emission measures, for these three spectra. It therefore appears that three of the four observations of V710 Tau N and V710 Tau S are reasonably well-characterized by single-component, low-temperature ($kT \sim$ 0.7-1.1 keV) plasma models (Table 2), whereas both the one- and two-component plasma models point to the presence of a dominant higher-temperature plasma ($kT \sim$ 2.1-2.6 keV) during the apparent flare on V710 Tau N captured in observation N1. 

High-resolution X-ray spectroscopy of cTTS and wTTS (e.g., \citet {kas02,kas04b}) show that X-ray emitting plasmas in T Tauri stars exhibit a range of temperatures and strong abundance anomalies, in particular, enhanced Ne. To investigate whether the T Tauri stars in the V710 Tau binary display such anomalies, we performed spectral fits with VMEKAL models, in which individual abundances can be varied. Given that the two-temperature MEKAL models had unconstrained emission measures for N2, S1, and S2, we attempted fits of one-temperature plasma VMEKAL models to these spectra, with Ne and Fe allowed to vary. All other metal abundances were set to 0.2 $Z_{\sun}$.  The single-component VMEKAL model slightly improves the fit for the S2 observation (Figure 4), however there is no improvement for S1 and N2 (Figure 5). A two-component VMEKAL model with an enhanced Ne abundance significantly improves the fit for the N1 observation, particularly in the $\sim1-1.5$ keV spectral region (Figure 3). We find both Ne and Fe to be slightly overabundant with respect to the nominal abundances, though the precise values are very uncertain due to poor photon counting statistics. Given the spectral resolution of the CCD data, we are unable to discern whether Ne IX or NeX, the dominant ions in the high-resolution spectra of accreting stars, is primarily responsible for the excess emission. However the abundances from the VMEKAL fitting suggest that the Ne/Fe ratio may be enhanced in the X-ray emitting plasma for V710 Tau N during the flare event.   

\section{Discussion}

\subsection{Spatially-resolved vs.\ unresolved X-ray spectroscopy of the V710 Tau
  binary}

The results of our spectral fitting for the
individual X-ray-emitting components of the V710 Tau wTTS/cTTS binary (\S
3.2) can be compared with previous results obtained for this system
from XMM-Newton EPIC observations, in which the binary is
unresolved. In the spectral analysis performed by \citet{fav03}, the
V710 Tau N/S combined X-ray spectrum was well-fit by an absorbed two-component plasma model, with $kT_1 = 0.63 \pm 0.05$ keV and $kT_2 = 1.24 \pm 0.10$ keV
and a coronal metal abundance Z = 0.22 $\pm$ 0.07 $Z_{\sun}$.  \citet{fav03}
further improved the model fit by varying the individual relative
metal abundances; this analysis suggested a Ne abundance of 0.98 $\pm$
0.45 $Z_{\sun}$, nearly a factor of five above the derived Fe abundance.  The results from \citet{fav03} are nearly identical to ours for both temperature and abundance values and the indication of enhanced Ne, which strongly suggests that the composite spectrum modeled by \citet{fav03} is dominated by the X-ray spectrum of V710 Tau N. 

\citet{giar06} fit the unresolved XMM-Newton EPIC low-resolution spectra of V710 Tau for eleven observations. Each observation was fit independently with an absorbed, single-component plasma model, resulting in best-fit values of $kT$
ranging from 0.41 $\pm$ 0.24 to 1.03 $\pm$ 0.06 keV and N$_{H}$
ranging from 0.10 $\pm$ 0.18 x 10$^{22}$ cm$^{-2}$ to 0.71 $\pm$ 0.27
x 10$^{22}$ cm$^{-2}$. 

Our results are consistent with both the \citet{fav03} and \citet{giar06} analyses, but further indicate {\it (a)} in quiescence, {\it both} components of the V710 Tau binary are relatively soft ($kT \sim$ 1.0 keV) X-ray sources and {\it (b)} harder ($kT \sim2.5$ keV) X-ray emission is only present during the apparent flare on V710 Tau N in 2004 December, based on both one- and two-component plasma model fits.  The latter result suggests that neither binary component was flaring during the 2000 September and 2004 March XMM observations analyzed by \citet{fav03} and \citet{giar06}. Indeed, these authors report no variability in the unresolved V710 Tau X-ray source during these XMM exposures. 

Our spectral fitting results for V710 Tau N are also
consistent with the findings of \citet{wol05}, who used two-component
plasma models to describe X-ray emission from strongly flaring T Tauri
stars in Orion. \citet{wol05} found that a ``hot'' plasma component
(median $kT \sim 7$ keV) in conjunction with a ``cool'' plasma
component ($kT \sim$0.7 - 0.9 keV) describes such sources well and,
furthermore, that the cool component is unaffected by the magnetic
activity that presumably causes the flaring. The same behavior is apparent
for V710 Tau N.

\subsection{Accretion signatures in the X-ray emission from V710 Tau?}

Evidence is growing that stellar mass accretion plays a role in the
production of some of the X-rays from some T Tauri stars. Some of this
evidence comes from analysis of TTS X-ray plasma temperatures.
\citet{kas02} and \citet{hue07} interpreted the presence of very low
temperature (0.15-0.25 keV) plasmas in high-resolution X-ray spectra of
TW Hya and the borderline wTTS/cTTS binary Hen3-600, respectively, as
an accretion signature. High-resolution X-ray spectra of other cTTS have
also revealed the presence of ``soft excesses'' associated with these
stars that are ascribed to accretion processes \citep{tel07b}. A
complementary effect is apparently seen in low-resolution (CCD) X-ray
spectroscopy. Specifically, CCD spectral studies of statistically
significant TTS samples indicate a trend wherein cTTS exhibit a
deficit of X-ray emission relative to wTTS; this deficit, which has also
been explained in terms of disk stripping (e.g., \citet{jar06,gre07}) or field complexity in the photosphere (e.g., \citet{joh07}), is evidently accompanied by an enhancement of high-$T_X$ emission in cTTS (Telleschi et al. 2007a and references therein). \citet{tel07a} interpret these results as indicating that the accreting material in cTTS cools active regions to low temperatures, decreasing the X-ray luminosity --- while increasing the average X-ray temperature --- of cTTS relative to wTTS.

Though one must be cautious in placing results for individual objects in the context of statistical trends, our spectral modeling results for the individual components of the V710 Tau system are consistent with the contrast in $T_X$ between cTTS and wTTS that is apparent in the XMM Taurus cloud survey data presented in \citet{tel07a}. Specifically, in both quiescent and flaring states, V710 Tau N consistently exhibits higher plasma temperatures than V710 Tau S (Table 2), and these results for $T_X$ for V710 Tau N and S fall within the respective ranges determined for cTTS and wTTS in Taurus (see Figure 14 of \citet{tel07a}). 

Our results, like those of \citet{fav03}, point to the likelihood of
enhanced Ne/Fe abundance ratios in the X-ray emitting plasma of V710
Tau N and, possibly, V710 Tau S. In the case of the cTTS TW Hya, such
an anomalous Ne/Fe ratio has been interpreted as an X-ray signature of
accretion (e.g., \citet{ste04}). However, high Ne/Fe ratios have also
been observed in coronally active stars, and for those stars the enhancements are usually attributed to
the inverse FIP (first ionization potential) effect (e.g.,
\citet{argi05}).  The plasma Ne/O ratio may be a more effective
diagnostic of accretion activity (\citet{dra05a}), but we are unable
to constrain this ratio for either component of the V710
Tau binary from our observations.

\section {Conclusions} 

We have presented the first spatially resolved X-ray observations
  of the wTTS/cTTS binary system V710 Tau. The two stars were cleanly
resolved by the Chandra/ACIS-S3 imager, allowing us to separately
analyze the X-rays emitted by each star. Both stars maintained
near-constant count rates within each observation. V710 Tau N was evidently undergoing a flare event during the December 2004 observation since its X-ray flux was a factor $\sim$5 larger in December 2004 than in April 2005.

The best-fit models to each component of the V710 Tau binary indicate
plasma temperatures in the range $kT_X \sim 0.7-1.1$ keV for V710 Tau
S in 2004 December\ and for both stars in 2005 April, and that the X-ray emitting plasma associated with V710 Tau N  was much hotter during the former exposure and marginally hotter during the latter exposure than the plasma associated with V710 Tau S. The 2004 December\ flare-state spectrum of V710 Tau N appears to require both a soft component and the addition of a second, higher-temperature plasma component ($kT_X \sim 2.5-3$ keV). There is also evidence for an enhanced Ne/Fe abundance ratio in the X-ray emitting plasma associated with V710 Tau N.

The spectral modeling results indicate that the individual components of this wTTS/cTTS binary system follow general trends apparent in recent statistical studies of cTTS and wTTS (e.g.,\citet{wol05,tel07a}). Specifically, the cTTS consistently exhibited somewhat harder X-ray emission than the wTTS, and the X-ray
spectrum of the cTTS hardened considerably during the December 2004
flare event. 

Hence, in the V710 Tau system, harder X-ray flux and possibly, an enhanced Ne/Fe abundance ratio appear to accompany enhanced stellar accretion. Additional, deeper Chandra exposures of V710 Tau and similar wTTS/cTTS binaries systems will better constrain the plausible sources of X-ray emission from low-mass, pre-MS stars. Deeper Chandra observations of V710 Tau N/S, as well as observations of similar ``mixed'' TTS binaries, are required to establish whether the characteristic plasma temperatures and abundances of cTTS and wTTS binary components differ consistently and significantly.

\bibliographystyle{apj} 
\bibliography{ms.bib}

\clearpage

\begin{deluxetable}{cllll}
\tabletypesize{\scriptsize}
\tablecaption{Hardness Ratios$^1$}
\tablewidth{0pt}
\tablecolumns{5}
\tablehead{
\colhead{Observation} & \colhead{count rate} & \colhead{HR1} & \colhead{HR2} & \colhead{HR3} \\
\colhead{}       &  \colhead{cnts/ks}   &  \colhead{}   & \colhead{}    &   \colhead{} \\
}
\startdata
V710 Tau N (Dec) & 74.3 $\pm$ 2.6 & -0.66 $\pm$ 0.04 & -0.64 $\pm$ 0.04 & -0.39 $\pm$ 0.08 \\
V710 Tau N (Apr) & 15.3 $\pm$ 1.2 & -0.83 $\pm$ 0.10 & -0.82 $\pm$ 0.10 & -0.56 $\pm$ 0.32 \\ 
V710 Tau S (Dec) & 23.8 $\pm$ 1.5 & -0.94 $\pm$ 0.09 & -0.91 $\pm$ 0.09 & -0.60 $\pm$ 0.36 \\ 
V710 Tau S (Apr) & 14.8 $\pm$ 1.2 & -0.96 $\pm$ 0.11 & -0.94 $\pm$ 0.11 & -1.00 $\pm$ 0.78 \\  
\enddata
\tablecomments {The hardness ratios were calculated as defined in \citet{get05}. HR3 of V710 Tau S in April is attributed to lack of counts in the hardest range. Errors for observations with low counts were estimated according to \citet{geh86}. }
\end{deluxetable}

\clearpage
\begin{deluxetable}{cccccclccc}
\tabletypesize{\scriptsize}
\setlength{\tabcolsep}{0.05in}
\tablecaption{Model parameters and results}
\tablewidth{0pt}
\tablecolumns{9}
\tablehead{
\colhead{Obs.} & \colhead{Model} & \colhead{F$_{X,abs.}^1$} & \colhead{L$_{X,intr.}^2$} & \colhead{kT$_{1}$} & \colhead{kT$_{2}$} & \colhead{EM$_1^2$} & \colhead{EM$_2^2$} & d.o.f. & \colhead {$\chi^2$} \\
\colhead{} & \colhead{} & \colhead {\tiny(10$^{-13}$ erg cm$^{-2}$s$^{-1}$)} & \colhead{log (erg s$^{-1}$)} &\colhead {(keV)} & \colhead{(keV)} & \colhead{(10$^{53}$ cm$^{-3}$)} &\colhead{(10$^{53}$ cm$^{-3}$)} & \colhead{} & \colhead{} \\ 
}
\startdata
N1 & 1T-MEKAL  & 3.16 - 3.67 & $29.9^{+1.07}_{-1.19}$ & 2.13 $\pm$ 0.18 & ... & 1.47 $\pm$ 0.12 & ... & 39 & 1.77 \\
N1 & 2T-MEKAL  & 3.11 - 3.75 & $29.9^{+1.31}_{-0.89}$ & 0.51 $\pm$ 0.27 & 2.64 $\pm$ 0.38 & 0.30 $\pm$ 0.18 & 1.25 $\pm$ 0.21 & 37 & 1.67 \\ 
N1 & 2T-VMEKAL & 3.09 - 3.80 & $29.9^{+1.23}_{-0.86}$ & 0.50 $\pm$ 0.19 & 3.15 $\pm$ 0.59 & 0.12 $\pm$ 0.12 & 1.40 $\pm$ 0.29 & 35 & 1.59\\ 
\tableline
N2 & 1T-MEKAL & 0.41 - 0.58 & $29.08^{+0.85}_{-0.72}$ & 1.01 $\pm$ 0.087 & ... & 0.31 $\pm$ 0.06 & ... & 6 &1.16  \\ 
\tableline
S1 & 1T-MEKAL & 0.69 - 0.95 & 29.28 $\pm$ 0.81 & 0.76 $\pm$ 0.054 & ... & 0.56 $\pm$ 0.08 & ... & 11 & 2.3 \\ 
\tableline
S2 & 1T-MEKAL & 0.48 - 0.65 &$29.13^{+0.84}_{-0.82}$ & 0.78 $\pm$ 0.065 & ... & 0.42 $\pm$ 0.07 & ... & 7 & 1.12 \\  
S2 & 1T-VMEKAL & 0.22 - 0.60 &$29.10^{+0.05}_{-0.38}$ & 0.58 $\pm$ 0.21 & ... & 0.61 $\pm$ 0.63 & ... & 5 & 0.89 \\
\enddata
\tablecomments {N$_{H}$ was set to the optically-derived value of 0.29 x 10$^{22}$ cm$^{-2}$. 1) The absorbed X-ray flux ranges are based on 68\% Bayesian confidence ranges. 2) L$_{X,intr.}$ and EM were computed assuming a distance of 140 pc to the V710 Tau system. }
\end{deluxetable}

\clearpage

\clearpage

\clearpage

\begin{figure}
\epsscale{.80}
\begin{center}
\begin{tabular}{cc}
\multicolumn{2}{c}{\epsfig{file=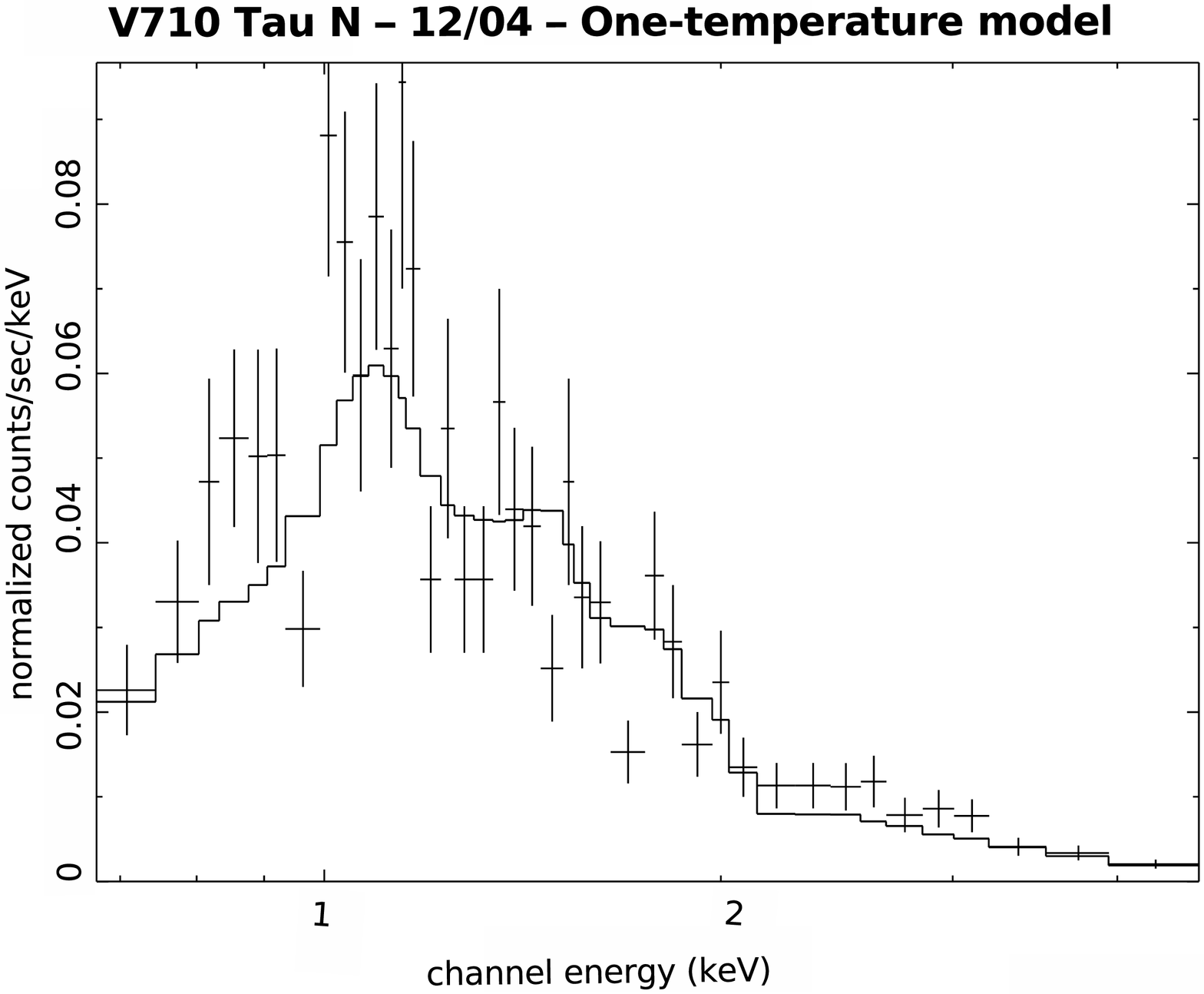,angle=270, width=3.2in}} \\
\epsfig{file=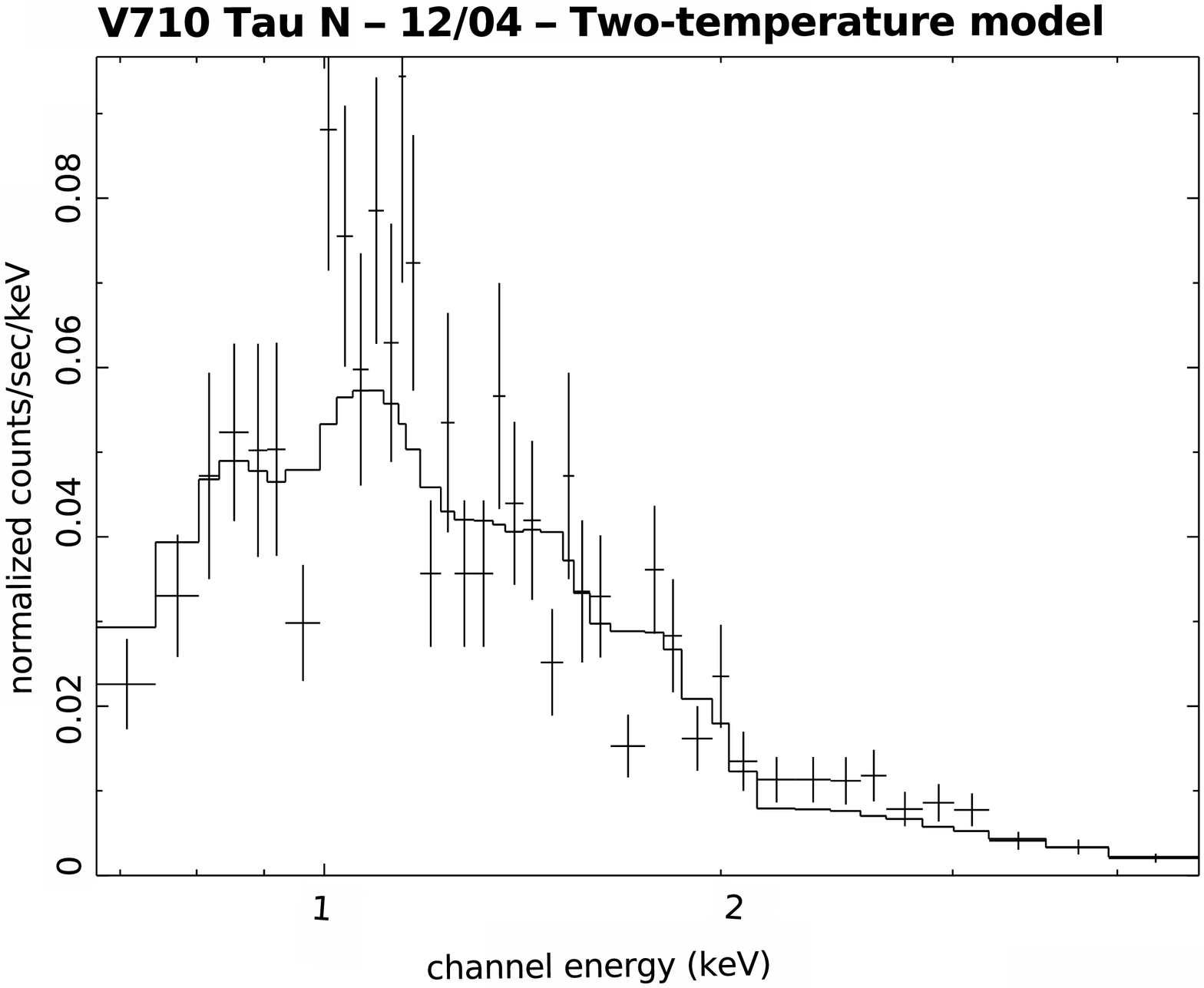,angle=270,width=3.2in} &
\epsfig{file=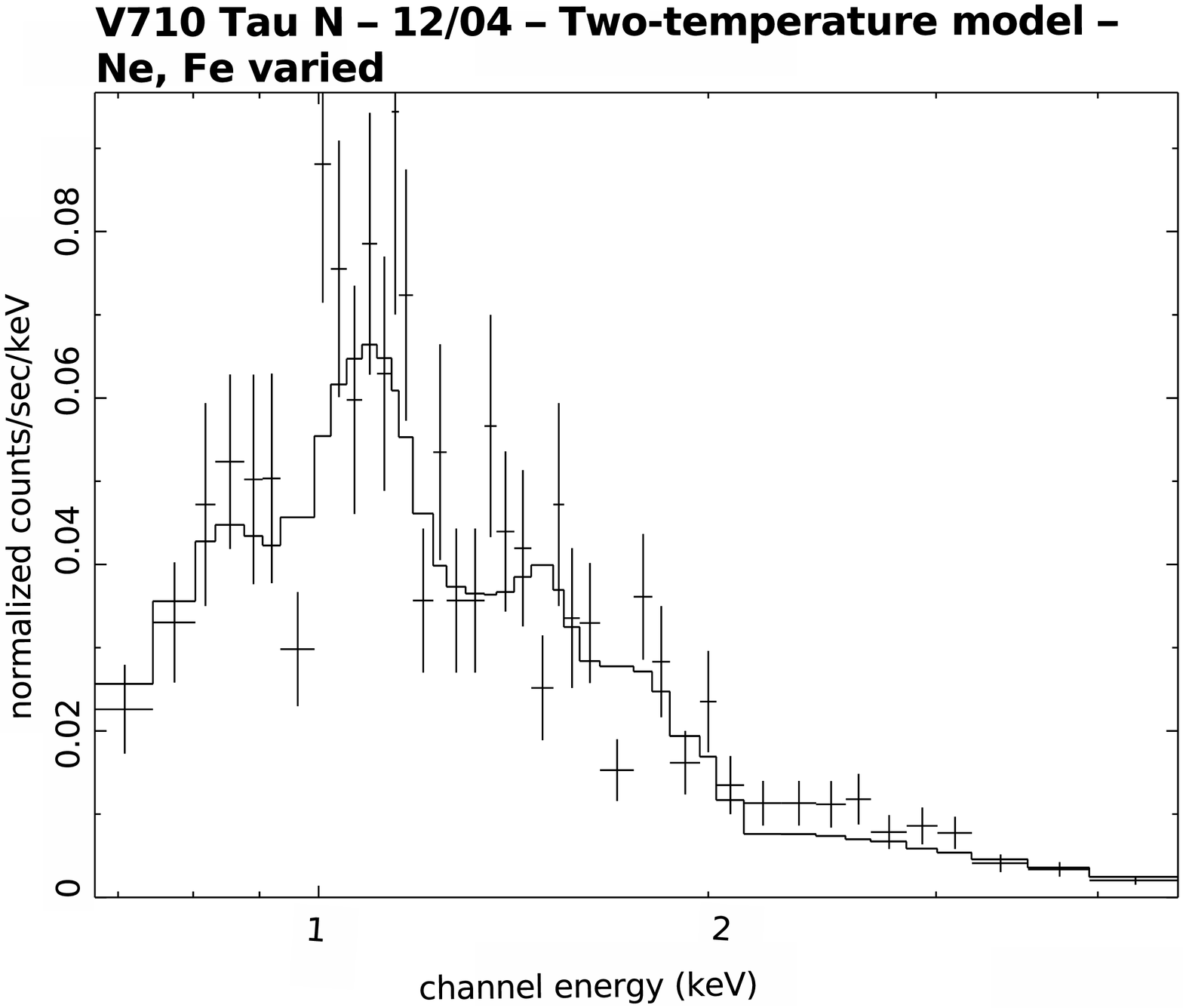,angle=270,width=3.2in} \\
\end{tabular} 
\caption{Model fits for V710 Tau N, December 2004 (N1) observation.  The model fits are a one-temperature plasma WABS(MEKAL)(top), two-temperature WABS(MEKAL)(bottom, left), and two-temperature WABS(VMEKAL) (bottom, right). The corresponding model parameters are listed in Table 2. A two-temperature model (bottom panels) is needed to fit low-energy data points in the region $<$ 1.0 keV. Note the improved fit in the $\sim1.0-1.5$ keV range in the bottom right panel.  The WABS(VMEKAL) (bottom right) model has a Ne abundance of 0.72 $\pm$ 0.92 $Z_{\sun}$ and a Fe abundance of 0.54 $\pm$ 0.30 $Z_{\sun}$ (all other elements at Z = 0.2 $Z_{\sun}$), which slightly improves the fit in the region around 1 keV. \label{fig4}}
\end{center}
\end{figure}

\begin{figure}
\begin{center}
\begin{tabular}{cc}
\epsfig{file=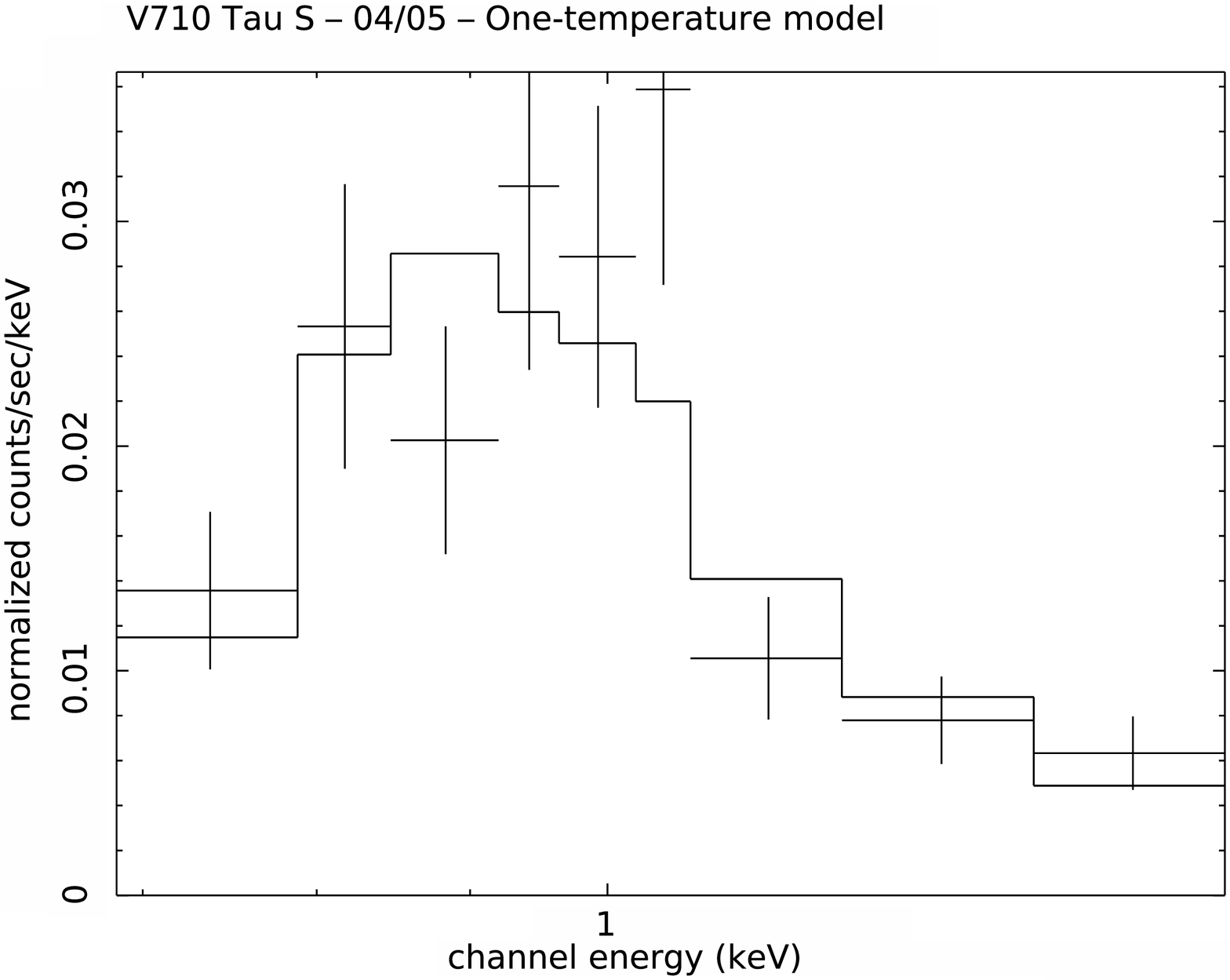,angle=270,width=3.5in}&
\epsfig{file=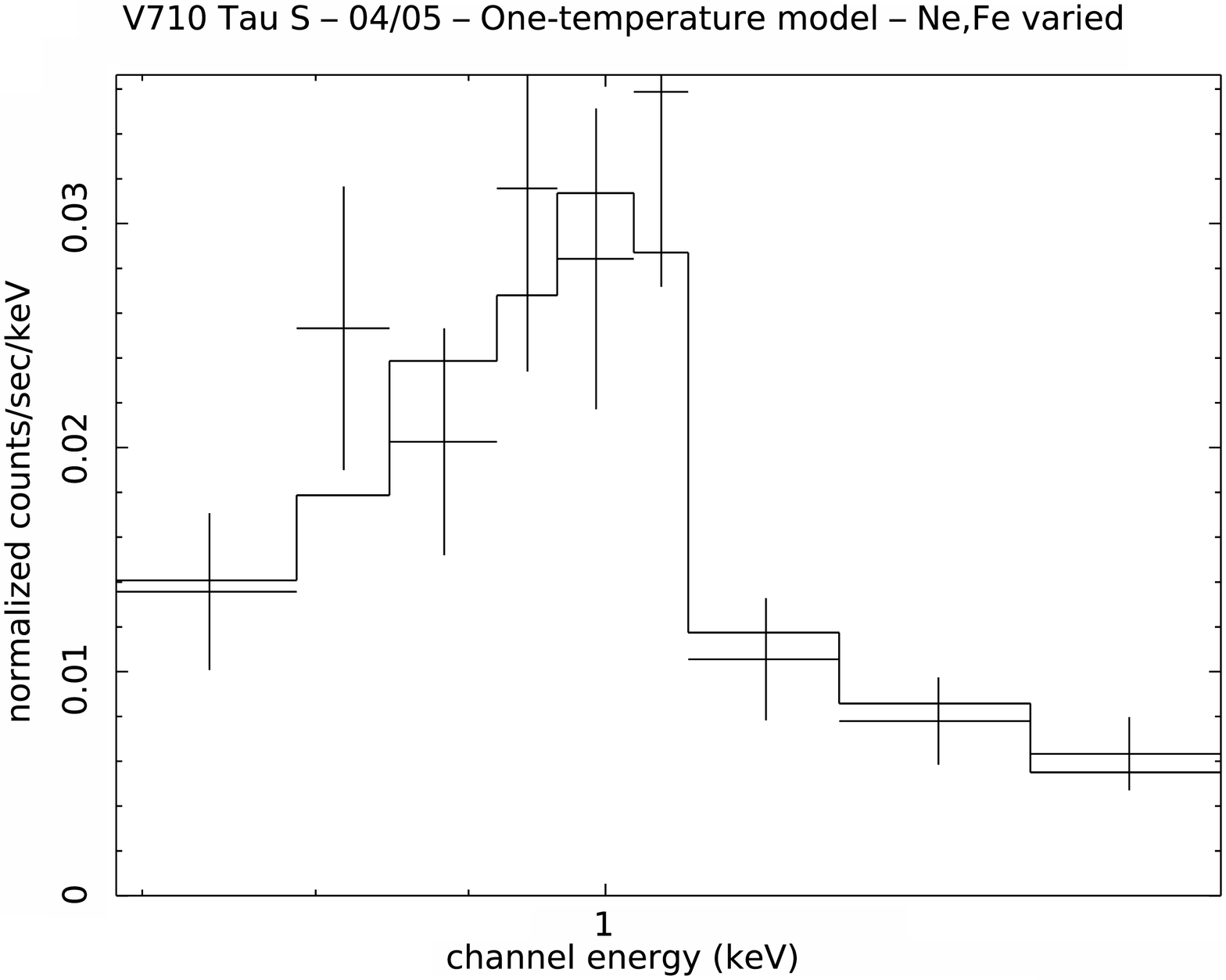,angle=270,width=3.5in}\\
\end{tabular}
\caption{V710 Tau S, April 2005 model fits. Single-temperature WABS(MEKAL) model with subsolar abundances (left) and a WABS(VMEKAL) model (right) with a Ne abundance of 0.78 $\pm$ 0.55 $Z_{\sun}$. The VMEKAL model provides a better fit at almost all energies.}
\end{center}
\end{figure}

\clearpage

\begin{figure}
\begin{center}
\begin{tabular}{cc}
\epsfig{file=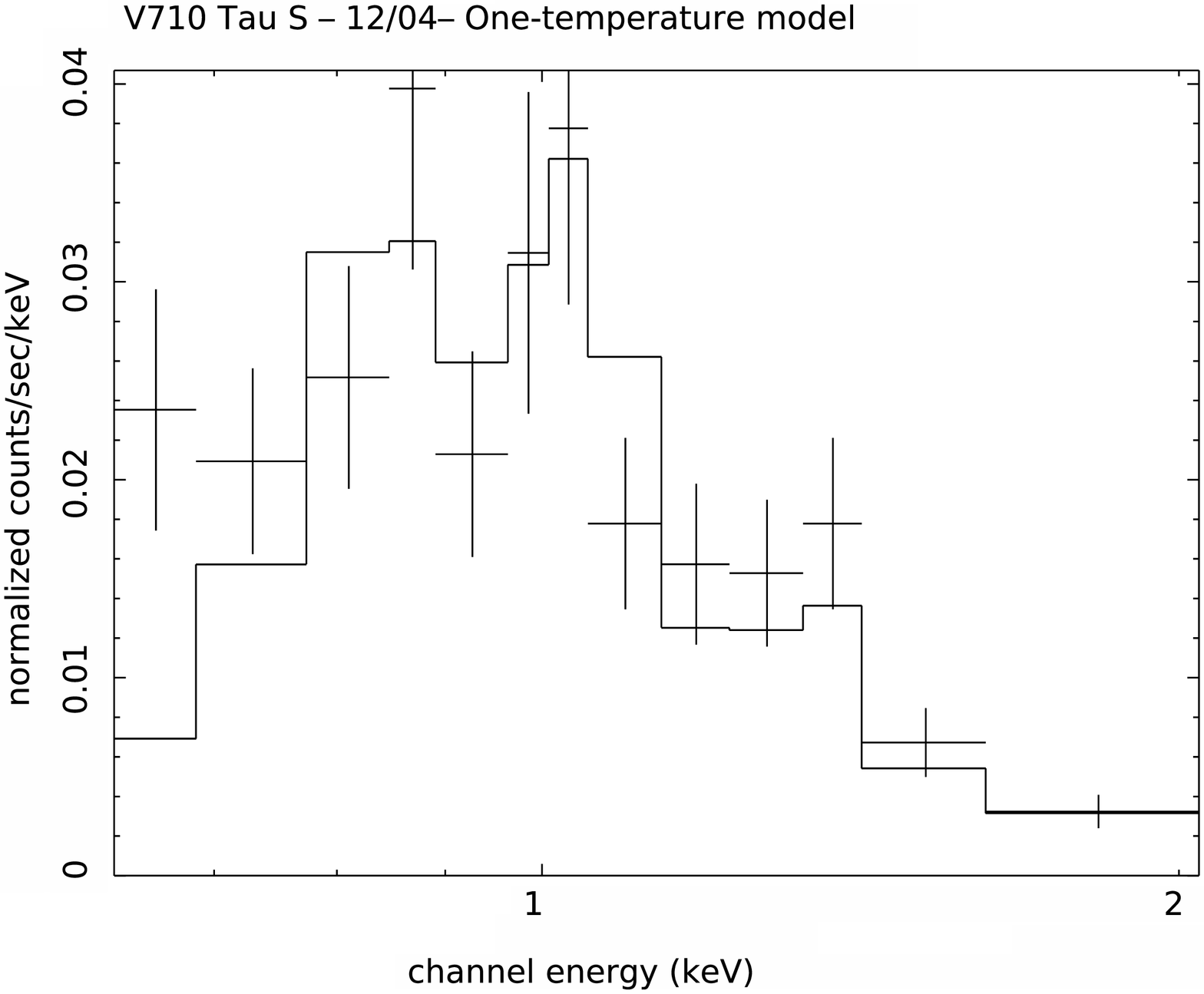,angle=270,width=3.5in} &
\epsfig{file=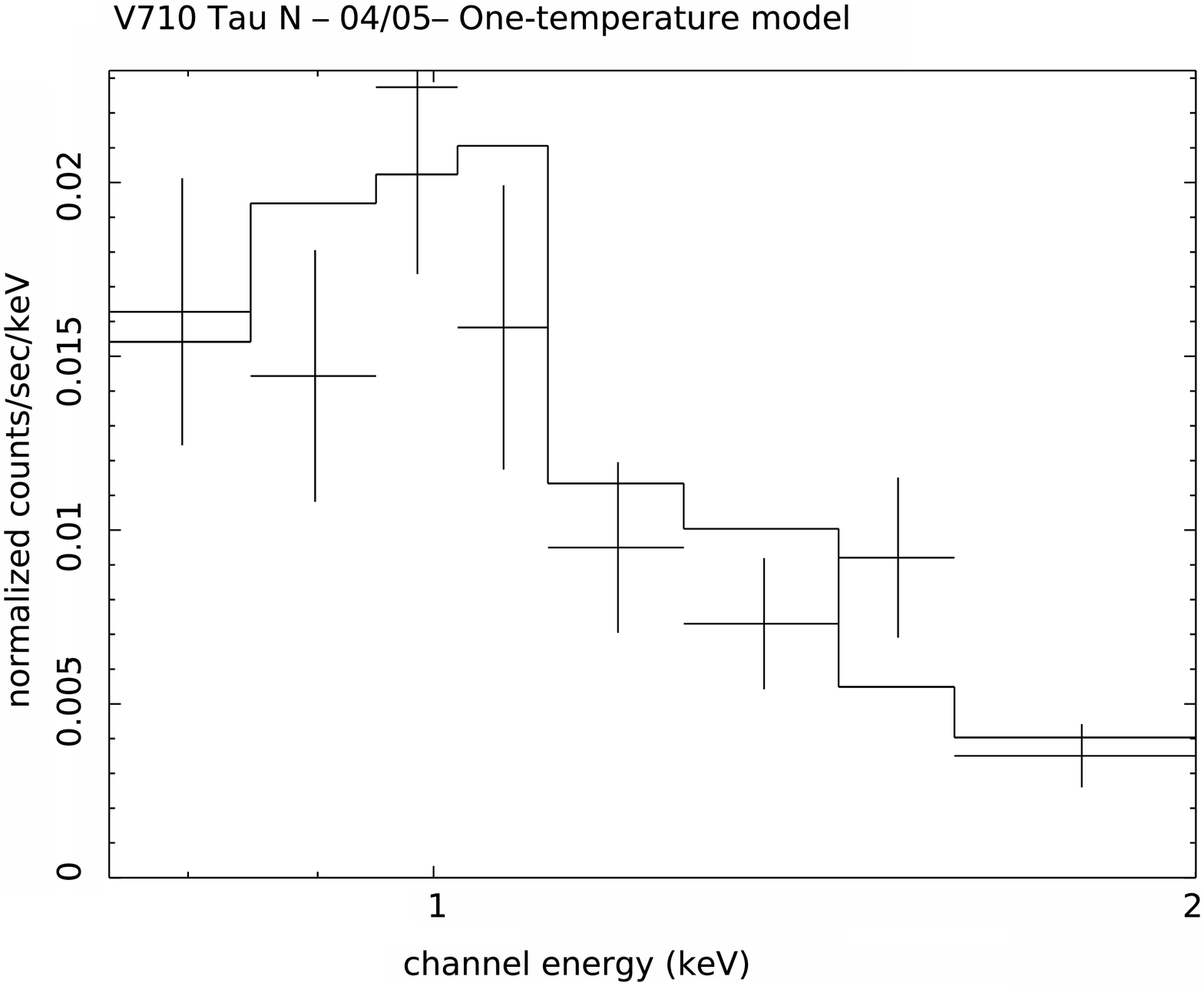,angle=270,width=3.5in}\\
\end{tabular}
\caption{Single-temperature WABS(MEKAL) model fits for V710 Tau S, December 2004 (left) and V710 Tau N, April 2005 (right).}
\end{center}
\end{figure}

\clearpage

\end{document}